\documentclass[twocolumn,eqsecnum,prb]{revtex4}
\usepackage[T1]{fontenc}
\usepackage{amsmath,amsbsy,amssymb,graphicx}

\def\beginABC{\begin{subequations}}
\def\endABC{\end{subequations}}
\begin{document}

\title{{\Large Coulomb Blockade in Graphene Nanodisks}}
\author{Motohiko Ezawa}
\affiliation{Department of Applied Physics, University of Tokyo, Hongo 7-3-1, 113-8656,
Japan }

\begin{abstract}
Graphene nanodisk is a graphene derivative with a closed edge. The trigonal
zigzag nanodisk with size $N$ has $N$-fold degenerated zero-energy states.
We investigate electron-electron interaction effects in the zero-energy
sector. We explicitely derive the direct and exchange interactions, which
are found to have no SU($N$) symmetry. Then, regarding a nanodisk as a
quantum dot with an internal degree of freedom, we analyze the nanodisk-lead
system consisting of a nanodisk and two leads. Employing the standard Green
function method, we reveal novel Coulomb blockade effects in the system. The
occupation number in the nanodisk exhibits a peculiar series of plateaux and
dips, reflecting a peculiar structure of energy spectrum of nanodisk without
SU($N$) symmetry. Dips are argued to emerge due to a Coulomb correlation
effect.
\end{abstract}

\maketitle


\address{{\normalsize Department of Applied Physics, University of Tokyo, Hongo
7-3-1, 113-8656, Japan }}

\section{Introduction}

Graphene-related materials has opened a new exciting field of research in
nanoelectronics.\cite{GraphExA,GraphExB,GraphExC} In particular, graphene
nanoribbons\cite%
{Fujita,EzawaPRB,Brey,Rojas,Son,Barone,Kim,Avouris,Xu,Ozyilmaz} have
attracted much attention due to a rich variety of band gaps, from metals to
wide-gap semiconductors. Another class of graphene derivatives are graphene
nanodisks.\cite{EzawaPhysica,EzawaDisk} They are nanometer-scale disk-like
materials which have closed edges. Graphene nanodisks can be constructed by
connecting several benzenes, some of which have already been manufactured by
soft-landing mass spectrometry.\cite{Rader,Kim,Berger} Nanodisks as well as
nanoribbons would be promising candidates of future electronic nanodevices.%
\cite{EzawaPhysica}

There are varieties of nanodisks, among which trigonal zigzag nanodisks are
prominent in their electronic property because there exist half-filled
zero-energy states. This novel property was revealed first based on
tight-binding model\cite{EzawaDisk} and confirmed subsequently by
first-principle calculations.\cite{Fernandez,Hod} We introduce the size
parameter $N$ for trigonal zigzag nanodisks as illustrated in Fig.\ref%
{FigNanodisk}(a). Then, there exists $N$-fold degenerated zero-energy states
in the trigonal zigzag nanodisk with size $N$. We have already argued\cite%
{EzawaDisk} that spins make a ferromagnetic order and that the relaxation
time is quite large even if the size $N$ is very small.

\begin{figure}[h]
\includegraphics[width=0.4\textwidth]{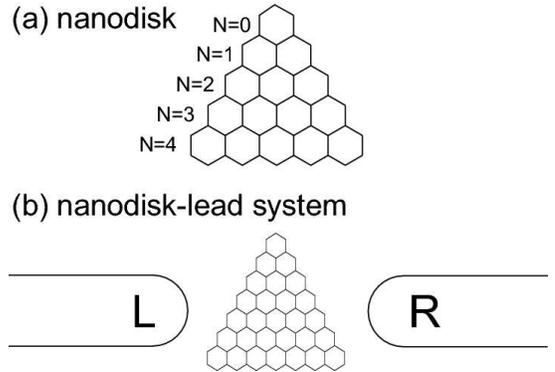}
\caption{(a) Geometric configuration of trigonal zigzag nanodisks. It is
convenient to introduce the size parameter $N$ in this way. The $0$-trigonal
nanodisk consists of a single Benzene, and so on. The number of carbon atoms
is given by $N_{\text{C}}=N^{2}+6N+6$. (b) Illustration of the nanodisk-lead
system. A nanodisk is connected to the right and left leads by tunneling
coupling. }
\label{FigNanodisk}
\end{figure}

In this paper we make an investigation of electron-electron interaction
effects in the zero-energy sector of the graphene nanodisk consisting of $N$
states. We derive explicitly the direct and exchange interactions, where
there is no SU($N$) symmetry. We estimate the spin stiffness. It is found to
be as large as a few hundred meV, which means that a nonodisk is indeed a
rigid ferromagnet. Then we analyze the nanodisk-lead system based on the
standard Green function method, where a nanodisk is connected to right and
left leads [Fig.\ref{FigNanodisk}(b)]. The analysis can be carried out quite
in the same way as in conventional quantum dots. The present system turns
out to be akin to a single dot system with an internal degree of freedom. We
find a series of Coulomb blockade peaks in the conductance as a function of
the chemical potential. Furthermore, the occupation number in the nanodisk
exhibits a peculiar series of plateaux and dips, reflecting a peculiar
structure of energy spectrum without SU($N$) symmetry.

This paper is organized as follows. In Sec. \ref{SecNanodisk}, we summarize
the basic nature of trigonal zigzag nanodisks. In Sec. \ref{Interaction}, we
study direct and exchange Coulomb interactions in the zero-energy sector. In
Sec. \ref{EfHamil}, formulating the nanodisk-lead system, we construct the
effective Hamiltonian near the half filling. In Sec. \ref{Blockade}, we
investigate the Coulomb blockade in the nanodisk-lead system, and calculate
numerically the occupation number in the nanodisk and the Coulomb blockade
peaks as a function of the chemical potential. Sec. \ref{Discuss} is devoted
to concluding remarks.

\section{Graphene Nanodisks}

\label{SecNanodisk}

Graphene nanodisks are graphene derivatives which have closed edges. The
Hamiltonian is defined by%
\begin{equation}
H=\sum_{i}\varepsilon _{i}c_{i}^{\dagger }c_{i}+\sum_{\left\langle
i,j\right\rangle }t_{ij}c_{i}^{\dagger }c_{j},  \label{HamilTB}
\end{equation}%
where $\varepsilon _{i}$ is the site energy, $t_{ij}$ is the transfer
energy, and $c_{i}^{\dagger }$ is the creation operator of the $\pi $
electron at the site $i$. The summation is taken over all nearest
neighboring sites $\left\langle i,j\right\rangle $. Owing to their
homogeneous geometrical configuration, we may take constant values for these
energies, $\varepsilon _{i}=\varepsilon _{\text{F}}$ and $t_{ij}=t$. We
choose $t=3$eV as a phenomenological parameter\cite{Saito}.

In a previous work\cite{EzawaDisk}, we have investigated the electronic and
magnetic properties of graphene nanodisks with various sizes and shapes in
quest of zero-energy states or equivalently metallic states. The emergence
of zero-energy states is surprisingly rare. Among typical nanodisks, only
trigonal zigzag nanodisks have degenerate zero-energy states and show
metallic ferromagnetism, where the degeneracy can be controlled arbitrarily
by designing the size.

It is convenient to introduce the size parameter $N$ for trigonal zigzag
nanodisks as in Fig.\ref{FigNanodisk}(a). The size-$N$ nanodisk has $N$-fold
degenerated zero-energy states\cite{EzawaDisk}, where the gap energy is as
large as a few eV. Hence it is a good approximation to investigate the
electron-electron interaction physics only in the zero-energy sector, by
projecting the system to the subspace made of those zero-energy states.

\begin{figure}[h]
\includegraphics[width=0.34\textwidth]{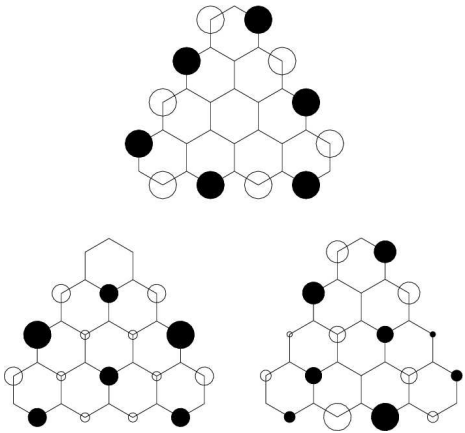}
\caption{The $3$ degenerate zero-energy states of the trigonal nanodisk with
size $N=3$. The solid (open) circle denotes that the amplitude $\protect%
\omega _{i}^{\protect\alpha }$ is positive (negative). The amplitude is
proportional to the radius of circle. It is seen that no electrons are
present in B sites.}
\label{FigWave}
\end{figure}

Let $|f_{\alpha }\rangle $ be the zero-energy state, $\alpha =1,2,\cdots ,N$%
. The wave function of the state $|f_{\alpha }\rangle $ is expanded as%
\begin{equation}
f_{\alpha }(\boldsymbol{x})=\sum_{i}\omega _{i}^{\alpha }\varphi _{i}(%
\boldsymbol{x}),  \label{EqA}
\end{equation}%
where $\varphi _{i}(\boldsymbol{x})$ is the Wannier function localized at
the site $i$, and $\omega _{i}^{\alpha }$ is the probability density to find
an electron there. The constraint condition is $\sum_{i}(\omega _{i}^{\alpha
})^{2}=1$.

Diagonalizing the Hamiltonian (\ref{HamilTB}), we are able to calculate
explicitly the amplitude $\omega _{i}^{\alpha }$ for zero-energy states in
the trigonal zigzag nanodisk. All of them are found to be real. As an
example we show them with size $N=3$ in Fig.\ref{FigWave}, where the solid
(open) circles denote the amplitude $\omega _{i}^{\alpha }$ is positive
(negative). The amplitude is proportional to the radius of circle. It is
seen that no electrons are present on B sites, where the graphene honeycomb
lattice is made of A and B sites.

\section{Electron-Electron Interactions}

\label{Interaction}

We take two states $|f_{\alpha }\rangle $ and $|f_{\beta }\rangle $, $\alpha
\neq \beta $, each of which can accommodate two electrons with up and down
spins at most. The two-state system is decomposed into the spin singlet $%
\chi _{\text{S}}$\ and the spin triplet $\chi _{\text{T}}$ with the
normalized wave functions,\label{ExchaWave}\beginABC%
\begin{align}
f_{\text{SS}}(\mathbf{x},\mathbf{x}^{\prime })& ={\frac{1}{\sqrt{2}}}\left(
f_{\alpha }(\mathbf{x})f_{\beta }(\mathbf{x}^{\prime })+f_{\alpha }(\mathbf{x%
}^{\prime })f_{\beta }(\mathbf{x})\right) \chi _{\text{S}}, \\
f_{\text{ST}}(\mathbf{x},\mathbf{x}^{\prime })& ={\frac{1}{\sqrt{2}}}\left(
f_{\alpha }(\mathbf{x})f_{\beta }(\mathbf{x}^{\prime })-f_{\alpha }(\mathbf{x%
}^{\prime })f_{\beta }(\mathbf{x})\right) \chi _{\text{T}}.
\end{align}%
\endABC The Coulomb energies are\beginABC%
\begin{align}
\langle f_{\text{SS}}|H_{\text{C}}|f_{\text{SS}}\rangle =& U_{\alpha \beta
}+J_{\alpha \beta }, \\
\langle f_{\text{ST}}|H_{\text{C}}|f_{\text{ST}}\rangle =& U_{\alpha \beta
}-J_{\alpha \beta },
\end{align}%
\endABC with\beginABC%
\begin{align}
U_{\alpha \beta }& =\int \!d^{3}xd^{3}y\;f_{\alpha }^{\ast }(\mathbf{x}%
)f_{\alpha }(\mathbf{x})V(\mathbf{x}-\mathbf{y})f_{\beta }^{\ast }(\mathbf{y}%
)f_{\beta }(\mathbf{y}), \\
J_{\alpha \beta }& =\int \!d^{3}xd^{3}y\;f_{\alpha }^{\ast }(\mathbf{x}%
)f_{\beta }(\mathbf{x})V(\mathbf{x}-\mathbf{y})f_{\beta }^{\ast }(\mathbf{y}%
)f_{\alpha }(\mathbf{y}),
\end{align}%
\label{EqX}\endABC where $V(\mathbf{x}-\mathbf{x}^{\prime })$ is the Coulomb
potential; $f_{\alpha }^{\ast }(\mathbf{x})f_{\alpha }(\mathbf{x})$ is the
electron number density in the state $|f_{\alpha }\rangle $, and $f_{\alpha
}^{\ast }(\mathbf{x})f_{\beta }(\mathbf{x})$ is the overlap of the wave
functions associated with the two states $|f_{\alpha }\rangle $ and $%
|f_{\beta }\rangle $. They are interpreted as the direct and exchange
energies.

Applying the above argument to the many-state system, the effective
Hamiltonian is derived as 
\begin{align}
H_{\text{D}}^{\text{eff}}=& \sum_{\alpha \geq \beta }U_{\alpha \beta
}n\left( \alpha \right) n\left( \beta \right)  \notag \\
& -\frac{1}{2}\sum_{\alpha >\beta }J_{\alpha \beta }[4\mathbf{S}(\alpha
)\cdot \mathbf{S}(\beta )+n\left( \alpha \right) n\left( \beta \right) ],
\label{HamilFerro}
\end{align}%
where $n\left( \alpha \right) $ is the number operator and $\mathbf{S}%
(\alpha )$ is the spin operator,%
\begin{equation}
n\left( \alpha \right) =d_{\sigma }^{\dag }(\alpha )d_{\sigma }(\alpha
),\qquad \mathbf{S}(\alpha )=\frac{1}{2}d_{\sigma }^{\dag }(\alpha )\mathbf{%
\tau }_{\sigma \sigma ^{\prime }}d_{\sigma ^{\prime }}(\alpha ),
\end{equation}%
with $d_{\sigma }(\alpha )$ the annihilation operator of electron with spin $%
\sigma =\uparrow ,\downarrow $ in the state $|f_{\alpha }\rangle $: $\mathbf{%
\tau }$ is the Pauli matrix. Note that we have included the on-state Coulomb
term $U_{\alpha \alpha }n\left( \alpha \right) n\left( \alpha \right) $ in
the effective Hamiltonian (\ref{HamilFerro}).

We expand $U_{\alpha \beta }$ and $J_{\alpha \beta }$ in terms of the
Wannier functions, \beginABC%
\begin{align}
U_{\alpha \beta }=& \sum_{s}\omega _{i}^{\alpha }\omega _{j}^{\alpha }\omega
_{k}^{\beta }\omega _{l}^{\beta }\int \!d^{3}xd^{3}y\;  \notag \\
& \times \varphi _{i}^{\ast }(\boldsymbol{x})\varphi _{j}(\boldsymbol{x})V(%
\mathbf{x}-\mathbf{y})\varphi _{k}^{\ast }(\boldsymbol{y})\varphi _{l}(%
\boldsymbol{y}), \\
J_{\alpha \beta }=& \sum_{s}\omega _{i}^{\alpha }\omega _{j}^{\alpha }\omega
_{k}^{\beta }\omega _{l}^{\beta }\int \!d^{3}xd^{3}y\;  \notag \\
& \times \varphi _{i}^{\ast }(\boldsymbol{x})\varphi _{j}(\boldsymbol{y})V(%
\mathbf{x}-\mathbf{y})\varphi _{k}^{\ast }(\boldsymbol{y})\varphi _{l}(%
\boldsymbol{x}).
\end{align}%
\endABC The dominant contributions come from the on-site Coulomb terms with $%
i=j=k=l$ both for the direct and exchange energies. We thus obtain%
\begin{equation}
U_{\alpha \beta }\simeq J_{\alpha \beta }\simeq {U}\sum_{i}(\omega
_{i}^{\alpha }\omega _{i}^{\beta })^{2},  \label{EqY}
\end{equation}%
with%
\begin{equation}
U\equiv \int \!d^{3}xd^{3}y\;\varphi _{i}^{\ast }(\boldsymbol{x})\varphi
_{i}(\boldsymbol{x})V(\mathbf{x}-\mathbf{y})\varphi _{i}^{\ast }(\boldsymbol{%
y})\varphi _{i}(\boldsymbol{y}).
\end{equation}%
The nearest neighbor Coulomb interaction vanishes since there are no
electrons in the nearest neighboring sites: See Fig.\ref{FigWave}.

We have numerically calculated $U_{\alpha \beta }$ for nanodisks with
several size $N$. For the case of $N=2$,%
\begin{equation*}
U_{11}=U_{22}=0.145U,\quad U_{12}=0.0482U,
\end{equation*}%
which gives $U_{11}/U_{12}=3$. For the case of $N=3$,%
\begin{align*}
U_{11}=& U_{22}=0.0944U,\quad U_{12}=0.0315U,\quad \\
U_{33}=& 0.0833U,\quad U_{13}=U_{23}=0.0655U,
\end{align*}%
which gives $U_{11}/U_{12}=3$, $U_{13}/U_{12}\simeq 2$. For the case of $N=4$%
,%
\begin{align*}
U_{11}=& 0.0556U,\quad U_{12}=0.0228U,\quad U_{13}=U_{14}=0.0435U, \\
U_{22}=& 0.0768U,\quad U_{23}=U_{24}=0.0478U, \\
U_{33}=& U_{44}=0.0934U,\quad U_{34}=0.0311U,
\end{align*}%
which gives $U_{33}/U_{34}=3$, $U_{13}/U_{12}\simeq 2$.

We make some remarkable observations. First, the exchange energy is as large
as the direct energy, which is the order of a few hundred meV. Thus the spin
stiffness $J_{\alpha \beta }$ is quite large, implying that nanodisks are
rigid ferromagnets. Furthermore, as we have seen numerically, all $J_{\alpha
\beta }$ are the same order of magnitude for any pair of $\alpha $ and $%
\beta $, implying that the SU($N$) symmetry is broken but not so strongly in
the Hamiltonian (\ref{HamilFerro}). Hence, the zero-energy sector is
described by the SU(N) Heisenberg model as a rough approximation. These
facts confirm our previous result\cite{EzawaDisk} that the relaxation time
of the ferromagnetic-like spin polarization is quite large even if the size
of trigonal zigzag nanodisks is very small.

\section{Nanodisk-Lead System}

\label{EfHamil}

We investigate the Coulomb blockade of nanodisks in a similar way to the
case of conventional quantum dots. We consider the system comprised of
nanodisks with right and left connected leads [Fig.\ref{FigNanodisk}(b)].
The Hamiltonian of the system is written as 
\begin{equation}
H=H_{\text{D}}+H_{\text{L}}+H_{\text{T}},  \label{TotalHamil}
\end{equation}%
where $H_{\text{D}}$ is given by (\ref{HamilFerro}), and\beginABC%
\begin{align}
H_{\text{L}}=& \sum_{k\sigma }\varepsilon \left( k\right) \left( c_{k\sigma
}^{\text{R}\dagger }c_{k\sigma }^{\text{R}}+c_{k\sigma }^{\text{L}\dagger
}c_{k\sigma }^{\text{L}}\right) , \\
H_{\text{T}}=& t_{\text{L}}\sum_{k\sigma }\sum_{\alpha }\left( c_{k\sigma }^{%
\text{L}\dagger }d_{\sigma }(\alpha )+d_{\sigma }^{\dagger }(\alpha
)c_{k\sigma }^{\text{L}}\right)  \notag \\
& +t_{\text{R}}\sum_{k\sigma }\sum_{\alpha }\left( c_{k\sigma }^{\text{R}%
\dagger }d_{\sigma }(\alpha )+d_{\sigma }^{\dagger }(\alpha )c_{k\sigma }^{%
\text{R}}\right) .  \label{HamilT}
\end{align}%
\endABC The Hamiltonian $H_{\text{L}}$ describes a noninteracting electron
gas in the leads with $\varepsilon \left( k\right) =\hbar ^{2}\mathbf{k}%
^{2}/2m$, while $H_{\text{T}}$ the tunneling interaction between the leads
and the nanodisk with $t_{\text{L(R)}}$ the tunneling coupling constant: We
have assumed that the spin does not flip in the tunneling process.

It is convenient to make the transformation%
\begin{equation}
\left( 
\begin{array}{c}
c_{k\sigma }^{\text{e}} \\ 
c_{k\sigma }^{\text{o}}%
\end{array}%
\right) =\frac{1}{\tilde{t}}\left( 
\begin{array}{cc}
t_{\text{L}}^{\ast } & t_{\text{R}}^{\ast } \\ 
-t_{\text{R}} & t_{\text{L}}%
\end{array}%
\right) \left( 
\begin{array}{c}
c_{k\sigma }^{\text{L}} \\ 
c_{k\sigma }^{\text{R}}%
\end{array}%
\right)
\end{equation}%
with%
\begin{equation}
\tilde{t}=\sqrt{\left\vert t_{L}\right\vert ^{2}+\left\vert t_{R}\right\vert
^{2}},
\end{equation}%
so that the right and left leads are combined into the "even" and "odd"
leads. The lead Hamiltonian $H_{\text{L}}$ is invariant under above
transformation,%
\begin{equation}
H_{\text{L}}=\sum_{k\sigma }\varepsilon \left( k\right) \left( c_{k\sigma }^{%
\text{e}\dagger }c_{k\sigma }^{\text{e}}+c_{k\sigma }^{\text{o}\dagger
}c_{k\sigma }^{\text{o}}\right) ,
\end{equation}%
but the transfer Hamiltonian is considerably simplified,%
\begin{equation}
H_{\text{T}}=\tilde{t}\sum_{k\sigma }\sum_{\alpha }\left( c_{k\sigma }^{%
\text{e}\dagger }d_{\sigma }(\alpha )+d_{\sigma }^{\dagger }(\alpha
)c_{k\sigma }^{\text{e}}\right) .
\end{equation}%
It looks as if the tunneling occurs only between the "even" lead and the
nanodisk.

In the case of $N=1$, the system is reduced to the simple Anderson model of
quantum dot%
\begin{equation}
H_{\text{D}}=\sum U_{11}n_{\uparrow }n_{\downarrow },
\end{equation}%
and hence the physics is well known\cite{Anderson}. However, for a general
case, it is hard to study the electronic properties of the system because
the interaction strengths $U_{\alpha \beta }$ take complicated values.

Another simple example is given by the system at the half-filling with $%
N\gtrsim 2$. As we have argued, the ground state is ferromagnet and the
relaxation time is very large even though $N$ is small. Based on this fact
we simplify the Hamiltonian (\ref{HamilFerro}) by assuming that the nanodisk
is a rigid ferromagnet with the spin direction $\downarrow $. Then, only
electrons with spin $\uparrow $ are dynamical within the nanodisk. We
explore the electric properties of such a system.

The nanodisk Hamiltonian is described only by the electrons with the spin $%
\uparrow $,%
\begin{equation}
H_{\text{D}}^{\prime }=\sum_{\alpha }\varepsilon _{\alpha }n_{\alpha
\uparrow }+\sum_{\alpha >\beta }U_{\alpha \beta }n_{\alpha \uparrow
}n_{\beta \uparrow },  \label{HamilNanoA}
\end{equation}%
where%
\begin{equation}
\varepsilon _{\alpha }=\sum_{\beta }U_{\alpha \beta }.
\end{equation}%
Here, $U_{\alpha \beta }$ stands for the Coulomb energy between the added
up-spin electron in the state $\alpha $ and the background down-spin
electron in the state $\beta $. Thus, $\varepsilon _{\alpha }$ is the
potential energy increased when one electron is added to the state $\alpha $.

\begin{figure}[h]
\includegraphics[width=0.4\textwidth]{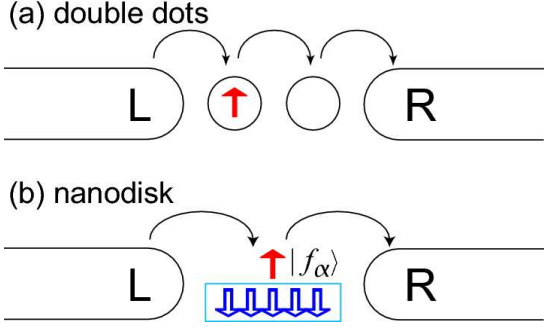}
\caption{{}(Color online) (a) Illustration of the standard double-dot
system. There is intra-hopping between dots. (b) Illustration of the nanodisk-lead system, where all down-spin electrons
are filled up in the nanodisk. An up-spin electron tunnels from the left lead to one of the
states $|f_{\protect\alpha }\rangle $ and then to the right lead. }
\label{FigFerroDisk}
\end{figure}

We first consider the case where one electron is tunnelled from the lead
into the nanodisk. Then it is enough to consider only the first term in the
Hamiltonian (\ref{HamilNanoA}). We have numerically calculated $\varepsilon
_{\alpha }$ for some nanodisks. For $N=2$,%
\begin{equation}
\varepsilon _{1}=\varepsilon _{2}=0.193U.
\end{equation}%
For $N=3$,%
\begin{equation}
\varepsilon _{1}=\varepsilon _{2}=0.191U,\quad \varepsilon _{3}=0.214U.
\end{equation}%
For $N=4$,%
\begin{equation}
\varepsilon _{1}=0.165U,\quad \varepsilon _{2}=0.195U,\quad \varepsilon
_{3}=\varepsilon _{4}=0.216U.
\end{equation}

This Hamiltonian looks similar to that of the $N$-dot system. However, there
exists a crucial difference. On one hand, in the ordinary $N$-dot system, an
electron hops from one dot to another dot [Fig.\ref{FigFerroDisk}(a)]. On
the other hand, in our nanodisk system, the index $\alpha $ of the
Hamiltonian runs over the $N$-fold degenerated states and not over the
sites. According to the Hamiltonian (\ref{HamilT}), an electron does not hop
from one state to another state [Fig.\ref{FigFerroDisk}(b)]. Hence, it is
more appropriate to regard our nanodisk as a one-dot system with an internal
degree of freedom. This fact simplifies the analysis considerably.

\section{Coulomb Blockade in Nanodisk}

\label{Blockade}

We investigate the Coulomb blockade in the nanodisk-lead system. It has been
argued that the conductance is given by the formula\cite{Meir},%
\begin{equation}
G=\frac{2e^{2}}{\hbar }\sum_{\alpha }\int d\varepsilon \frac{\Gamma ^{\text{L%
}}\Gamma ^{\text{R}}}{\Gamma ^{\text{L}}+\Gamma ^{\text{R}}}A\left( \alpha
,\varepsilon \right) \left( -\frac{\partial n_{F}\left( \varepsilon \right) 
}{\partial \varepsilon }\right) ,  \label{FormuCondu}
\end{equation}%
where $A\left( \alpha ,\varepsilon \right) $ is the spectral function of the
nanodisk system, $n_{\text{F}}\left( \varepsilon \right) $ is the Fermi
distribution function. The coupling strength $\Gamma ^{i}$ is given by%
\begin{equation}
\pi \rho \left( \varepsilon \right) \left\vert t_{i}\right\vert ^{2}=\left\{ 
\begin{array}{cc}
\Gamma ^{i} & \text{for }\left\vert \varepsilon \right\vert <D \\ 
0 & \text{for }\left\vert \varepsilon \right\vert >D%
\end{array}%
\right. ,
\end{equation}%
where $\rho \left( \varepsilon \right) $ is the density of state in the lead
at energy $\varepsilon $, which is almost a constant within the band limit, $%
\left\vert \varepsilon \right\vert <D$.

The spectral function is given by the retarded Green function as%
\begin{equation}
A\left( \alpha ,\varepsilon \right) =-\frac{1}{\pi }\text{Im}G^{\text{R}%
}\left( \alpha ,\varepsilon \right) ,
\end{equation}%
with%
\begin{equation}
G^{\text{R}}\left( \alpha ,t-t^{\prime }\right) =-i\theta \left( t-t^{\prime
}\right) \left\langle \left\{ d_{\alpha }\left( t\right) ,d_{\alpha
}^{\dagger }\left( t^{\prime }\right) \right\} \right\rangle .
\label{BasicGreen}
\end{equation}%
It exhibits the energy spectrum of the states indexed by the quantum number $%
\alpha $. At the zero-temperature, the formula (\ref{FormuCondu}) is reduced
to%
\begin{equation}
G=\frac{2e^{2}}{\hbar }\frac{\Gamma ^{\text{L}}\Gamma ^{\text{R}}}{\Gamma ^{%
\text{L}}+\Gamma ^{\text{R}}}\sum_{\alpha }A\left( \alpha ,\mu \right)
\end{equation}%
with $\mu $ the chemical potential, where the conductance is simply
proportional to the sum of the spectral density. This formula can be
interpreted as follows. If there is a state at the Fermi energy, an electron
can tunnel from the lead to the nanodisk via the state by ballistic
transport. If not, an electron needs to go via high energy states with
finite gap.

It is a straightforward task to derive the spectral function by employing
the standard technique\cite{Lacroix}, as we describe in the appendix. Here
let us summarize the result. The spectral function is simply given by the
sum of Lorentzians,%
\begin{align}
A\left( \alpha ,\varepsilon \right) =\frac{1}{\pi }& \sum_{\beta \neq \alpha
}\left[ \frac{\left( 1-\left\langle n_{\beta }\right\rangle \right) \Gamma }{%
\left( \varepsilon -\varepsilon _{\alpha }\right) ^{2}+\Gamma ^{2}}\right. 
\notag \\
& \qquad +\left. \frac{\left\langle n_{\beta }\right\rangle \Gamma }{\left(
\varepsilon -\varepsilon _{\alpha }-U_{\alpha \beta }\right) ^{2}+\Gamma ^{2}%
}\right] .  \label{Spectral}
\end{align}%
It has peaks at the energies $\varepsilon _{\alpha }$ and $\varepsilon
_{\alpha }+U_{\alpha \beta }$ with broadening $\Gamma $. The height of the
peak at energies $\varepsilon _{\alpha }$ is proportional to $1-\left\langle
n_{\beta }\right\rangle $, and at energies $\varepsilon _{\alpha }+U_{\alpha
\beta }$ is proportional to $\left\langle n_{\beta }\right\rangle $. The
occupation number $\left\langle n_{\beta }\right\rangle $ should be
determined self-consistently by solving%
\begin{equation}
(N-1)\left\langle n_{\alpha }\right\rangle =\int d\varepsilon \;n_{F}\left(
\varepsilon \right) A\left( \alpha ,\varepsilon \right) .
\end{equation}%
We insert the spectral function (\ref{Spectral}) into the above equation,
and obtain the linear equation for $\left\langle n_{\alpha }\right\rangle $,%
\begin{align}
& (N-1)\left\langle n_{\alpha }\right\rangle  \notag \\
=& (N-1)\left( \frac{1}{2}-\frac{1}{\pi }\tan ^{-1}\frac{\varepsilon
_{\alpha }}{\Gamma }\right)  \notag \\
& +\sum_{\beta \neq \alpha }\frac{\left\langle n_{\beta }\right\rangle }{\pi 
}\left( \tan ^{-1}\frac{\varepsilon _{\alpha }}{\Gamma }-\tan ^{-1}\frac{%
\varepsilon _{\alpha }+U_{\alpha \beta }}{\Gamma }\right) .
\label{Occupation}
\end{align}%
Our task is to solve this equation to determine $\left\langle n_{\alpha
}\right\rangle $.

In order to get an overview of the result, we first consider the atomic
limit, which is the zero tunneling-coupling limit ($\tilde{t}\rightarrow 0$%
), implying that $\Gamma \rightarrow 0$. The equations become very simple
and we can obtain analytical results. This is indeed a good approximation
because $\Gamma $ is very small. In this limit the spectral function (\ref%
{Spectral}) becomes%
\begin{align}
A\left( \alpha ,\varepsilon \right) =& \sum_{\beta \neq \alpha }[\left(
1-\left\langle n_{\beta }\right\rangle \right) \delta \left( \varepsilon
-\varepsilon _{\alpha }\right)   \notag \\
& \qquad +\left\langle n_{\beta }\right\rangle \delta \left( \varepsilon
-\varepsilon _{\alpha }-U_{\alpha \beta }\right) ],
\end{align}%
and the self-consistent equation (\ref{Occupation}) becomes%
\begin{align}
(N-1)\left\langle n_{\alpha }\right\rangle =& \sum_{\beta \neq \alpha
}[\left( 1-\left\langle n_{\beta }\right\rangle \right) \theta \left( \mu
-\varepsilon _{\alpha }\right)   \notag \\
& \qquad +\left\langle n_{\beta }\right\rangle \theta \left( \mu
-\varepsilon _{\alpha }-U_{\alpha \beta }\right) ],
\end{align}%
where $\theta \left( x\right) $ is the step function: $\theta \left(
x\right) =0$ for $x<0$ and $\theta \left( x\right) =1$ for $x\geq 0$. It is
easy to see that $\left\langle n_{\alpha }\right\rangle =0$ if $\mu
<\varepsilon _{a}$, and that $\left\langle n_{\alpha }\right\rangle =1$ if $%
\mu >\varepsilon _{\alpha }+$max$_{\beta }[U_{\alpha \beta }]$. Though the
occupation numbers $\left\langle n_{\alpha }\right\rangle $ are nontrivial
in the other region, it is straightforward to determine them.

\begin{figure}[t]
\includegraphics[width=0.5\textwidth]{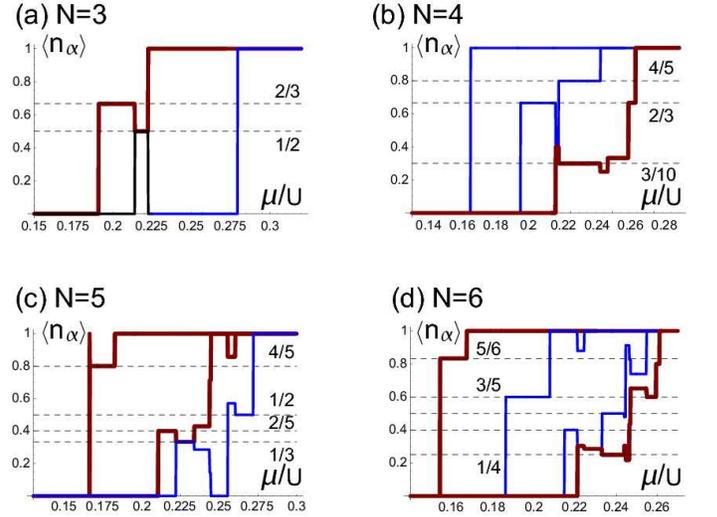}
\caption{{}(Color online) The occupation numbers $\left\langle n_{\protect%
\alpha }\right\rangle $ against the\ chemical potential in the case of $%
\Gamma ^{L}=\Gamma ^{R}=0$. (a) for $N=3$, (b) for $N=4$, (c) for $N=5$, and
(d) for $N=6$. Bold red lines denote the doubly degenerated occupancy, while
thin blue lines denote non-degenerated occupancy.}
\label{FigAtomic}
\end{figure}

We have shown $A\left( \alpha ,\varepsilon \right) $ and $\left\langle
n_{\alpha }\right\rangle $ as a function of the chemical potential $\mu $ in
Fig.\ref{FigAtomic}. For the case of $N=3$, the first plateau emerges at $%
\varepsilon _{1}=\varepsilon _{2}<\mu <\varepsilon _{3}$ with $%
n_{1}=n_{2}=2/3$, and the second plateau emerges at $\varepsilon _{3}<\mu
<\varepsilon _{1}+U_{12}$ with $n_{1}=n_{2}=n_{3}=1/2$. For the case of $N=4$%
, the first plateau emerges at $\varepsilon _{2}<\mu <\varepsilon
_{3}=\varepsilon _{4}$ with $n_{2}=2/3$, the second plateau emerges at very
small region $\varepsilon _{3}=\varepsilon _{4}<\mu <\varepsilon _{2}+U_{12}$
with $n_{1}=n_{2}=n_{3}=2/5$, and the third plateau emerges at $\varepsilon
_{2}+U_{12}<\mu <\varepsilon _{2}+U_{23}$ with $n_{2}=4/5$, $n_{3}=3/10$,
and so on.

\begin{figure}[t]
\includegraphics[width=0.5\textwidth]{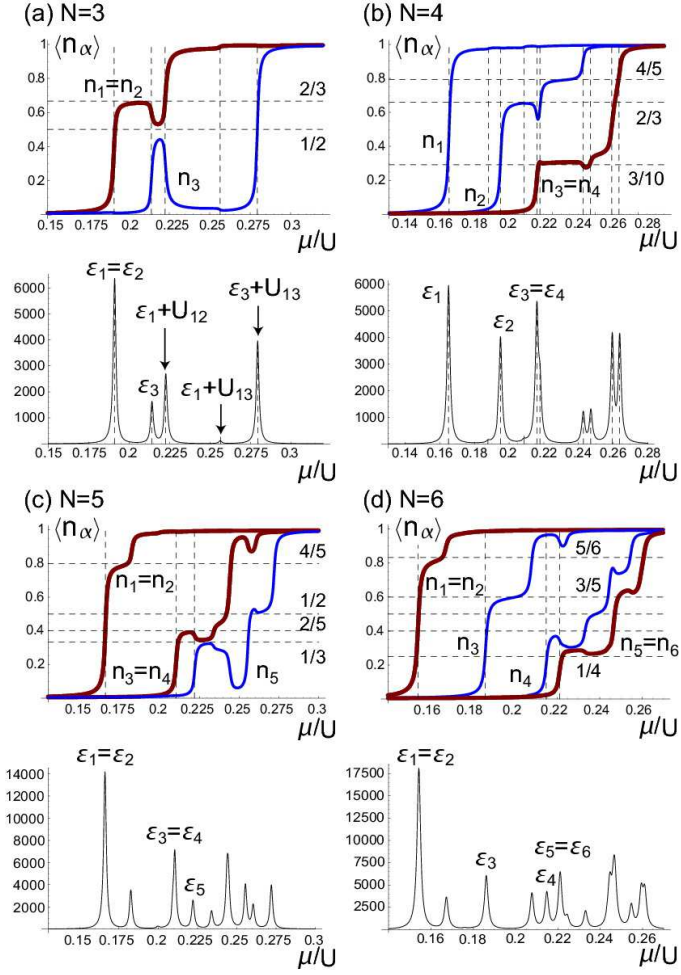}
\caption{{}(Color online) The occupation numbers $\left\langle n_{\protect%
\alpha }\right\rangle $ (upper figure) and the dimensionless conductance $G/(%
\frac{2e^{2}}{\hbar }\frac{\Gamma ^{L}\Gamma ^{R}}{\Gamma ^{L}+\Gamma ^{R}})$
(lower figure) against the\ chemical potential, where it is taken that $%
\Gamma ^{L}=\Gamma ^{R}=0.001$. (a) for $N=3$, (b) for $N=4$, (c) for $N=5$,
and (d) for $N=6$. Bold red lines denote the doubly degenerated occupancy,
while thin blue lines denote non-degenerated occupancy.}
\label{FigConduc2}
\end{figure}

We go on to study the case with a finite value of the tunneling coupling. We
have calculated numerically the occupation number $\left\langle n_{\alpha
}\right\rangle $ by solving (\ref{Occupation}), and then the spectral
function $A\left( \alpha ,\varepsilon \right) $ by returning it to (\ref%
{Spectral}). We show the occupation number and the conductance at the
zero-temperature in Fig.\ref{FigConduc2}.

Coulomb blockade peaks appear at $\mu =\varepsilon _{\alpha }$ and $\mu
=\varepsilon _{\alpha }+U_{\alpha \beta }$, where new channels open. There
are many peaks because the symmetry of the interactions is low, for example, 
$U_{11}\neq U_{12}$. There would be only two peaks if the SU($N$) symmetry
were exact. There are no peaks corresponding to the energy $\varepsilon
_{\alpha }+U_{\alpha \beta }+U_{\alpha \gamma }$ because we have neglected
the double occupancy in the nanodisk state.

It is intriguing that the occupancy $\left\langle n_{\alpha }\right\rangle $
shows a peculiar behavior containing dips as a function of the chemical
potential $\mu $. Note that, in the usual Coulomb blockade, a monotonic
behavior without dips are found in $\left\langle n_{\alpha }\right\rangle $
against the chemical potential. The occupancy makes a drastic change at the
energy of the Coulomb blockade peak.

Let us explain the origin of dips by taking an instance of the $N=3$
nanodisk [Fig.\ref{FigConduc2}(a)]. First, an electron enters the states $%
|f_{1}\rangle $ or $|f_{2}\rangle $ at $\mu =\varepsilon _{1}=\varepsilon
_{2}$, where all other up-spin states are empty, as results in the increase
of the occupancy $\left\langle n_{1}\right\rangle =\left\langle
n_{2}\right\rangle $. Similarly, an electron enters the state $|f_{3}\rangle 
$ at $\mu =\varepsilon _{3}$ together with the increase of $\left\langle
n_{3}\right\rangle $. However, this is accompanied with a dip in $%
\left\langle n_{1}\right\rangle =\left\langle n_{2}\right\rangle $ due to a
correlation effect: An electron in the state $|f_{3}\rangle $ prefers the
absence of electrons in the states $|f_{1}\rangle $ and $|f_{2}\rangle $
because of the Coulomb repulsion, thus decreasing the occupancy $%
\left\langle n_{1}\right\rangle =\left\langle n_{2}\right\rangle $. Next, an
electron enters the states $|f_{1}\rangle $ at $\mu =\varepsilon _{1}+U_{12}$
in the presence of an up-spin electron in the state $|f_{2}\rangle $, as
results in the increase of $\left\langle n_{1}\right\rangle $. In this way,
the occupancy $\left\langle n_{\alpha }\right\rangle $ shows a sudden
increase at $\varepsilon _{\alpha }$ and at $\varepsilon _{\alpha
}+U_{\alpha \beta }$, and a decrease at $\varepsilon _{\beta }$ and $%
\varepsilon _{\beta }+U_{\beta \gamma }$ for $\beta \neq \alpha $.

It is interesting that the peak at $\varepsilon _{1}+U_{13}$ is tiny in Fig.%
\ref{FigConduc2}(a). It is interpreted as follows. The peak occurs when an
electron enters the state $|f_{1}\rangle $ in the setting that the state $%
|f_{3}\rangle $ is already occupied by an up-spin electron. Though such a
setting is classically impossible because $\varepsilon _{1}<\varepsilon _{3}$%
, it can occur quantum-mechanically. Hence the process generates a tiny
peak. Similar peaks are observed at $\varepsilon _{1}+U_{12}$ and $%
\varepsilon _{1}+U_{13}$ in an instance of the $N=4$ nanodisk [Fig.\ref%
{FigConduc2}(b)].

In passing we note that the interaction strengths $\varepsilon _{\alpha }$
and $U_{\alpha \beta }$ can be determined experimentally by measuring
Coulomb blockade peaks.

\section{Conclusions}

\label{Discuss}

We have investigated correlation effects in the zero-energy sector of
graphene nanodisks consisting of $N$ states. We have derived explicitly the
direct and exchange interactions. It is found that there is no SU($N$)
symmetry. We have revealed a novel series of dips in the occupation number
on nanodisk together with Coulomb blockade peaks in the conductance as a
function of the chemical potential. Dips are argued to emerge due to a
Coulomb correlation effect. It is interesting that the interaction strengths 
$\varepsilon _{\alpha }$ and $U_{\alpha \beta }$ can be determined
experimentally by measuring these peaks.

We have estimated the spin stiffness in nanodisks at the half filling, which
is found to be as large as a few hundred meV. Thus there exists a strong
ferromagnetic coupling in the ground state of nanodisk. The relaxation time
is quite large even if the size is small, and the ground state can be
regarded as a rigid ferromagnet. Hence only electrons with spin opposite to
the direction of this ferromagnetic state can go through the nanodisk. We
may regard the nanodisk-lead system as the spin filter provided the electron
spin does not flip during the tunneling process between the nanodisk and
leads.

I am very much grateful to P. Kim and N. Nagaosa for many fruitful
discussions on the subject. The work was in part supported by Grants-in-Aid
for Scientific Research from Ministry of Education, Science, Sports and
Culture (Nos.070500000466).

\appendix

\section{Green's Function}

We analyze Green's function 
$G^{\text{R}}\left( \alpha ,t-t^{\prime }\right)$, which is defined by 
(\ref{BasicGreen}). Using the Heisenberg
equation of motion, $i\hbar \partial _{t}d_{\alpha }\left( t\right)
=[d_{\alpha }\left( t\right) ,H_{\text{D}}+H_{\text{L}}+H_{\text{T}}]$, we
find%
\begin{align}
& \left( i\hbar \partial _{t}-\varepsilon _{\alpha }\right) G^{\text{R}%
}\left( \alpha ,t-t^{\prime }\right)   \notag \\
=& \hbar \delta \left( t-t^{\prime }\right) +\sum_{\beta }U_{\alpha \beta
}G_{\beta }^{\text{R}}\left( \alpha ,t\right) +\sum_{k}\tilde{t}F^{\text{R}%
}\left( k,t-t^{\prime }\right) ,  \label{EqGd}
\end{align}%
where\beginABC%
\begin{align}
G^{\text{R}}\left( \alpha ,t-t^{\prime }\right) =& -i\theta \left(
t-t^{\prime }\right) \left\langle \left\{ d_{\alpha }\left( t\right)
,d_{\alpha }^{\dagger }\left( t^{\prime }\right) \right\} \right\rangle , \\
F^{\text{R}}\left( k,t-t^{\prime }\right) =& -i\theta \left( t-t^{\prime
}\right) \left\langle \left\{ c_{k}\left( t\right) ,d_{\alpha }^{\dagger
}\left( t^{\prime }\right) \right\} \right\rangle , \\
G_{\beta }^{\text{R}}\left( \alpha ,t-t^{\prime }\right) =& -i\theta \left(
t-t^{\prime }\right) \left\langle \left\{ n_{\beta }\left( t\right)
d_{\alpha }\left( t\right) ,d_{\alpha }^{\dagger }\left( t^{\prime }\right)
\right\} \right\rangle .
\end{align}%
\endABC Note that we have introduced the notation $F^{\text{R}}$ for the
correlation between the lead and the nanodisk.

We again use the Heisenberg equation of motion to calculate $\partial _{t}F^{%
\text{R}}\left( k,t-t^{\prime }\right) $,%
\begin{equation}
\left( i\hbar \partial _{t}-\varepsilon _{k}\right) F^{\text{R}}\left(
k,t-t^{\prime }\right) =\tilde{t}^{\ast }G^{\text{R}}\left( \alpha
,t-t^{\prime }\right) .  \label{EqC}
\end{equation}%
Making the Fourier transformations of (\ref{EqGd}) and (\ref{EqC}) and
combining them, we obtain%
\begin{equation}
\left( \hbar \omega -\varepsilon _{\alpha }-\Sigma ^{R}\left( \omega \right)
\right) G^{\text{R}}\left( \alpha ,\omega \right) =\hbar +\sum_{\beta
}U_{\alpha \beta }G_{\beta }^{\text{R}}\left( \alpha ,\omega \right) ,
\label{EqD}
\end{equation}%
where%
\begin{equation}
\Sigma ^{\text{R}}\left( \omega \right) =\sum_{k}\frac{\left\vert \tilde{t}%
\right\vert ^{2}}{\hbar \omega -\varepsilon _{k}+i\eta }\simeq -\frac{%
2\Gamma }{\pi }\ln \left\vert \frac{D+\hbar \omega }{D-\hbar \omega }%
\right\vert -i\Gamma 
\end{equation}%
is the self-energy, describing virtual electron tunneling between the
nanodisk and the lead. The real part gives a shift of energy, which is to be
used to define the new on-state energy 
\begin{equation*}
\varepsilon _{\alpha }^{\prime }=\varepsilon _{\alpha }+\text{Re}\Sigma ^{%
\text{R}}.
\end{equation*}%
However, since the energy shift is the order of $\varepsilon \Gamma /D$ and
small, we may neglect it. The imaginary part, which is independent of the
band width $D$, gives the width of the energy peak.

It is necessary to construct 
$G_{\beta }^{\text{R}}\left( \alpha ,t-t^{\prime}\right) $ 
to obtain Green's function $G^{\text{R}}\left( \alpha ,t-t^{\prime }\right) $ by (\ref{EqD}).
For this purpose we again use the Heisenberg equation of motion,
\begin{align}
i\hbar \partial _{t}G_{\beta }^{\text{R}}\left( \alpha ,t-t^{\prime }\right)
=& \hbar \delta \left( t-t^{\prime }\right) \left\langle \left\{ n_{\beta
}\left( t\right) d_{\alpha }\left( t\right) ,d_{\alpha }^{\dagger }\left(
t^{\prime }\right) \right\} \right\rangle   \notag \\
& \hspace{-0.85cm}-i\theta \left( t\right) \left\langle \left\{ -\left[
H,n_{\beta }\left( t\right) d_{\alpha }\left( t\right) \right] ,d_{\alpha
}^{\dagger }\left( t^{\prime }\right) \right\} \right\rangle .
\end{align}%
The commutator in the last term on the right-hand side results in the two
terms,%
\begin{equation}
\left[ H,n_{\beta }d_{\alpha }\right] =\left[ H_{D},n_{\beta }d_{\alpha }%
\right] +\left[ H_{T},n_{\beta }d_{\alpha }\right] ,
\end{equation}%
where\beginABC%
\begin{align}
\left[ H_{D},n_{\beta }d_{\alpha }\right] =& -\left( \varepsilon _{\alpha
}+U_{\alpha \beta }\right) n_{\beta }d_{\alpha }-\sum_{\beta \neq \gamma
}U_{\beta \gamma }n_{\gamma }n_{\beta }d_{\alpha },  \label{EqEa} \\
\left[ H_{T},n_{\beta }d_{\alpha }\right] =& -n_{\beta }\sum_{k}\tilde{t}%
c_{k}+\sum_{k}\tilde{t}\left( c_{k}^{\dagger }d_{\beta }-d_{\beta }^{\dagger
}c_{k}\right) d_{\alpha }.  \label{EqEb}
\end{align}%
\endABC Here we neglect higher-order Green functions\cite{Lacroix,Czycholl}.
First, we neglect the last term in (\ref{EqEa}), which corresponds to
neglect the double occupancy states of initial states. Next, we neglect the
last term in (\ref{EqEb}), which corresponds to neglect spin flips on the
nanodisk during the tunneling process. In this way we achieve at%
\begin{equation}
\left( i\hbar \partial _{t}-\varepsilon _{\alpha }-U_{\alpha \beta }\right)
G_{\beta }^{\text{R}}\left( \alpha ,t\right) =\hbar \delta \left( t\right)
\left\langle n_{\beta }\right\rangle +\sum \tilde{t}F_{\beta }^{\text{R}%
}\left( k,t\right) ,  \label{EqGR}
\end{equation}%
where we have introducee a new function,%
\begin{equation}
F_{\beta }^{\text{R}}\left( k,t-t^{\prime }\right) =-i\theta \left(
t-t^{\prime }\right) \left\langle \left\{ n_{\beta }\left( t\right)
c_{k}\left( t\right) ,d_{\alpha }^{\dagger }\left( t^{\prime }\right)
\right\} \right\rangle .
\end{equation}%
Thus the equation of motion for $G_{\beta }^{\text{R}}\left( \alpha
,t\right) $ produces a new function.

It is necessary to analyze the equation of motion of $F_{\beta }^{\text{R}%
}\left( \alpha ,t\right) $, which leads to the following commutators,%
\beginABC%
\begin{align}
\left[ H,n_{\beta }c_{k}\right] =& -\varepsilon _{k}c_{k}+\left[ H_{T},c_{k}%
\right] , \\
\left[ H_{T},n_{\beta }c_{k}\right] =& -\tilde{t}n_{\beta }c_{k}+\sum \tilde{%
t}\left( c_{k}^{\dagger }d_{\beta }-d_{\beta }^{\dagger }c_{k}\right) c_{k}.
\end{align}%
\endABC Again we neglect the last term, which corresponds to neglect a
spin-flip process\cite{Lacroix,Czycholl}. In this way we achieve at%
\begin{equation}
\left( i\hbar \partial _{t}-\varepsilon _{k}\right) F_{\beta }^{\text{R}%
}\left( k,t\right) =\tilde{t}G_{\beta }^{\text{R}}\left( \alpha ,t\right) ,
\label{EqFR}
\end{equation}%
which no longer produces a new function.

Now, (\ref{EqGR}) and (\ref{EqFR}) make a closed set of the equation of
motions, which we can solve easily. Making their Fourier transformations, we
obtain%
\begin{equation}
\left( \hbar \omega -\varepsilon _{\alpha }-U_{\alpha \beta }-\Sigma ^{\text{%
R}}\left( \omega \right) \right) G_{\beta }^{\text{R}}\left( \alpha ,\omega
\right) =\hbar \left\langle n_{\beta }\right\rangle .  \label{EqF}
\end{equation}%
It follows from (\ref{EqD}) and (\ref{EqF}) that%
\begin{align}
G^{\text{R}}\left( \alpha ,\omega \right) =\hbar & \sum_{\beta \neq \alpha }%
\left[ \frac{1-\left\langle n_{\beta }\right\rangle }{\hbar \omega
-\varepsilon _{\alpha }-\Sigma ^{R}\left( \omega \right) }\right.  \notag \\
& \left. +\frac{\left\langle n_{\beta }\right\rangle }{\hbar \omega
-\varepsilon _{\alpha }-U_{\alpha \beta }-\Sigma ^{R}\left( \omega \right) }%
\right] ,
\end{align}%
with $\varepsilon =\hbar \omega $. Hence the spectral function is given by
the sum of Lorentzians,%
\begin{align}
A\left( \alpha ,\varepsilon \right) =\frac{1}{\pi }& \sum_{\beta \neq \alpha
}\left[ \frac{\left( 1-\left\langle n_{\beta }\right\rangle \right) \Gamma }{%
\left( \varepsilon -\varepsilon _{\alpha }\right) ^{2}+\Gamma ^{2}}\right. 
\notag \\
& +\left. \frac{\left\langle n_{\beta }\right\rangle \Gamma }{\left(
\varepsilon -\varepsilon _{\alpha }-U_{\alpha \beta }\right) ^{2}+\Gamma ^{2}%
}\right] ,
\end{align}%
which is (\ref{Spectral}) in text.

\end{document}